\documentclass[11pt]{article}
\usepackage{amsmath}
\usepackage{amssymb}
\usepackage{amsthm,amsxtra}
\usepackage{epsf}
\usepackage{eepic}
\newlength{\mytopmargin}
\newlength{\myleftmargin}
\newtheorem{prop}{Proposition}
\newcommand{\zz}{\mathbb Z}

\newcommand\psymmU{%
\begin{picture}(1,1)(0,0)%
\allinethickness{0.5pt}%
\path(0,0)(0,1)(1,1)(1,0)(0,0)%
\end{picture}}
\newcommand\psymmUU{%
\begin{picture}(1,1)(0,0)%
\allinethickness{0.5pt}%
\path(0,0)(0,1)(1,1)(1,0)(0,0)%
\put(0.5,0.5){\makebox(0,0){$\cdot$}}%
\end{picture}}
\newcommand\psymmO{%
\begin{picture}(1,1)(0,0)%
\allinethickness{0.5pt}%
\path(0,0)(0,1)(1,1)(1,0)(0,0)%
\path(0,0)(1,1)%
\end{picture}}
\newcommand\psymmS{%
\begin{picture}(1,1)(0,0)%
\allinethickness{0.5pt}%
\path(0,0)(0,1)(1,1)(1,0)(0,0)%
\path(1,0)(0,1)%
\end{picture}}
\newcommand\psymmu{%
\begin{picture}(1,1)(0,0)%
\allinethickness{0.5pt}%
\path(0,0)(0,1)(1,1)(1,0)(0,0)%
\path(0,0)(1,1)%
\path(0,1)(1,0)%
\end{picture}}

\newbox\tsymmUbox
\newbox\tsymmUUbox
\newbox\tsymmObox
\newbox\tsymmSbox
\newbox\tsymmubox
\setbox\tsymmUbox =\hbox{\kern0.75pt\setlength{\unitlength}{6pt}\psymmU \kern0.75pt}

\setbox\tsymmUUbox=\hbox{\kern0.75pt\setlength{\unitlength}{6pt}\psymmUU\kern0.75pt}
\setbox\tsymmObox =\hbox{\kern0.75pt\setlength{\unitlength}{6pt}\psymmO \kern0.75pt}
\setbox\tsymmSbox =\hbox{\kern0.75pt\setlength{\unitlength}{6pt}\psymmS \kern0.75pt}
\setbox\tsymmubox =\hbox{\kern0.75pt\setlength{\unitlength}{6pt}\psymmu \kern0.75pt}

\newbox\symmUbox
\newbox\symmUUbox
\newbox\symmObox
\newbox\symmSbox
\newbox\symmubox
\setbox\symmUbox =\hbox{\kern0.75pt\setlength{\unitlength}{4.5pt}\psymmU \kern0.75pt}
\setbox\symmUUbox=\hbox{\kern0.75pt\setlength{\unitlength}{4.5pt}\psymmUU\kern0.75pt}
\setbox\symmObox =\hbox{\kern0.75pt\setlength{\unitlength}{4.5pt}\psymmO \kern0.75pt}
\setbox\symmSbox =\hbox{\kern0.75pt\setlength{\unitlength}{4.5pt}\psymmS \kern0.75pt}
\setbox\symmubox =\hbox{\kern0.75pt\setlength{\unitlength}{4.5pt}\psymmu \kern0.75pt}
\def\symmU{{\copy\symmUbox}}
\def\symmUU{{\copy\symmUUbox}}
\def\symmO{{\copy\symmObox}}
\def\symmS{{\copy\symmSbox}}
\def\symmu{{\copy\symmubox}}

\begin{document}

\vspace{4cm}
\noindent
{\bf Symmetrized models of last passage percolation and
non-intersecting lattice paths} 

\vspace{5mm}
\vspace{5mm}
\noindent
Peter J.~Forrester
 and Eric M.~Rains

\noindent
${}^*$Department of Mathematics and Statistics,
University of Melbourne, \\
Victoria 3010, Australia ; \\
${}^\dagger$ Department of Mathematics, 
University of California, Davis, CA 95616, USA
\small
\begin{quote}
It has been shown that the last passage time in certain symmetrized models
of directed percolation can be written in terms of averages over random
matrices from the classical groups $U(l)$, $Sp(2l)$ and $O(l)$. We present
a theory of such results based on non-intersecting lattice paths, and integration
techniques familiar from the theory of random matrices.
Detailed derivations of probabilities relating to two further
symmetrizations are also given.
\end{quote}

\section{Introduction}
\begin{figure}
\epsfxsize=4.5cm
\centerline{\epsfbox{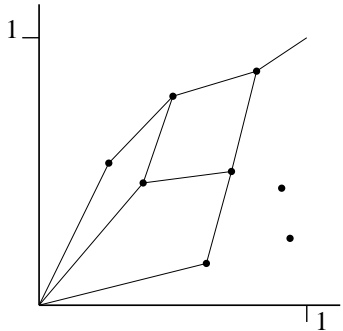}}
\caption{\label{bp.f15a} Eight points in the unit square, and the extended
directed paths of maximum length. Since the number of segments in these
paths equals four, here $l^\symmU_n = 3$. 
}
\end{figure}

There are a number of striking results linking models of stochastic
processes to
to random matrix
theory (for a recent work of this type see \cite{BFS07a}; for reviews see
\cite{Fo03,Sp05}). 
As an easy to explain example, consider
the following last passage percolation problem due to Hammersley. In the
unit square mark in points uniformly at random according to a Poisson
rate with intensity $\lambda$ (thus the probability the square contains
$N$ points is equal to ${\lambda^N \over N!} e^{-\lambda }$).
Join points by straight line segments with the requirement that the segments
have positive slope and form a continuous path, and extend this path to
begin at $(0,0)$ and finish at $(1,1)$. Define the
length of the extended path as the number of points it contains, and
denote by 
$l^\symmU = l^\symmU(\lambda)$ the stochastic variable specifying the
maximum of the lengths of all possible extended paths
(see Figure \ref{bp.f15a}). 
Then it is known from the work of Gessel \cite{Ge90}
and 
Rains \cite{Ra98} (see also \cite{Fo01}) that
\begin{equation}\label{5.1}
{\rm Pr}(l^\symmU \le l) = \Big 
\langle \prod_{j=1}^l e^{\sqrt{\lambda} \cos \theta_j}
\Big \rangle_{U(l)}
\end{equation}
where the average is with respect to the eigenvalue probability density
function (p.d.f.) of random matrices chosen uniformly at random from the group
$U(l)$. The latter has the explicit form
\begin{equation}\label{5.1a}
{1 \over (2 \pi)^l l!} \prod_{1 \le j < k \le l}
| e^{i \theta_k} - e^{i \theta_j} |^2.
\end{equation}
The formula (\ref{5.1}) is central to the proof by Baik, Deift and
Johansson \cite{BDJ98} giving the limiting scaled distribution of the
longest   increasing subsequence of a random permutation.

In a substantial work Baik and Rains \cite{BR01a} have considered symmetrized
generalizations of the Hammersley process, and have shown that the
cumulative probability for the analogue of the stochastic variable
$l^\symmU $ can be written as an average over the classical groups
$Sp(2l)$, $O(l)$ (for two particular symmetries), or $U(2l)$,
$U(l)\oplus U(l)$ (for two other symmetries). Moreover, random matrix
formulas were also given \cite{BR01}
for lattice generalizations of these processes
(in the case of the original Hammersley process the lattice generalization
is referred to as the Johansson model \cite{Jo99a}).
The proofs of \cite{BR01a} make sophisticated use of symmetric function
theory, and have independent interest in that discipline. On the
other hand it is reasonable to suggest that many researchers interested
in directed percolation and growth processes will lack the necessary
background in symmetric function theory to fully appreciate these proofs.
This has motivated us to seek alternative derivations which to a large
extent avoid the heavy use of symmetric function theory called upon
in \cite{BR01a}. Instead our proofs make use of a non-intersecting path
picture of the Johansson growth model \cite{Jo02,FR02b}, 
an extension of this picture to a last passage percolation model with
Bernoulli random variables,
together with
techniques familiar from the theory of random matrices
(the applicability of such techniques have been foreshadowed in Section
6 of \cite{BR01a}). Another
essential ingredient from \cite{BR01a} is the use of bijections from the
theory of Young tableaux (see e.g.~\cite{Fu97}). Here our presentation differs
from that of \cite{BR01a} only in that we give more detail. 

Our task then is to derive, from a non-intersecting paths picture, formulas
known from \cite{BR01a} giving cumulative probabilities of a suitable last
passage percolation variable for generalizations of the
Hammersely process in terms of random matrix averages. As already
mentioned, one such generalization is the Johansson model. In Section 2 its
definition is recalled, as is its formulation in the non-intersecting
path picture. Formulas from the theory of non-intersecting paths are then
used to derive the analogue of (\ref{5.1}). 
An analogous discussion of a variant of this model, involving Bernoulli rather
than geometric random variables, is given in Section 3.
The four symmetrized versions
of the Johansson model are then treated separately in each of the subsequent
four sections.

\section{Johansson model and polynuclear growth}
\setcounter{equation}{0}
We begin by defining the last passage percolation model introduced
by Johansson \cite{Jo99a}. Consider the upper right quadrant square lattice
$\{ (i,j): \, i,j \in \zz^+ \}$. Associate with each lattice site $(i,j)$
a random non-negative integer variable $x_{i,j}$ chosen from the geometric
distribution with parameter $a_i b_j$ so that
\begin{equation}\label{2.1}
{\rm Pr}(x_{i,j} = k) = (1 - a_i b_j) (a_i b_j)^k.
\end{equation}
Denote by $(1,1){\rm u/rh}(n,n)$ a sequence of lattice paths starting at
$(1,1)$ and finishing at $(n,n)$ with each lattice point successively
connected by edges which are either directed upwards or 
horizontally to the right.
One defines the last passage time $L_n^{\symmU}$ say as the maximum of the
sum of the integer variables associated with these lattice points. Thus
\begin{equation}\label{3.0}
L_n^{\symmU} := {\rm max } \sum_{(1,1){\rm u/rh}(n,n)} x_{i,j}.
\end{equation}
According to \cite{BR01a}, for given parameters $\{a_i\}, \{b_j\}$ the
cumulative distribution can be written as a random matrix average
according to
\begin{equation}\label{3.1}
{\rm Pr}( L_n^{\symmU} \le l) =
\prod_{i,j=1}^n(1-a_i b_j)
\Big \langle \prod_{j=1}^n \prod_{k=1}^l (1 + a_j e^{-i \theta_k})
(1 + b_j e^{i \theta_k}) \Big \rangle_{U(l)}
\end{equation}
where the average over $U(l)$ refers to the probability density function
(\ref{5.1a}). We seek a derivation of  (\ref{3.1}), and analogous formulas
from \cite{BR01a} for symmetrized versions of the Johansson model,
within a non-intersecting paths representation of the model.

\begin{figure}
\epsfxsize=7cm
\centerline{\epsfbox{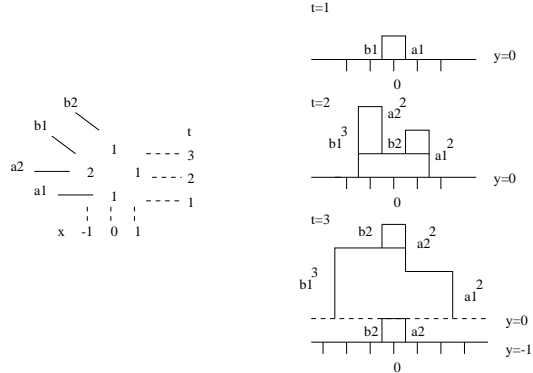}}
\caption{\label{bp.f15b} RSK correspondence in the polynuclear growth
model picture between a weighted non-negative integer matrix and a
pair of weighted non-intersecting lattice paths.
}
\end{figure}

Let us first revise how non-intersecting paths relate to the
Johansson model \cite{Jo02,FR02b}. 
This is done via a geometrical construction, equivalent
to the Robinson-Schensted-Knuth (RSK) correspondence from the theory of
Young tableaux \cite{Fu97}, which gives a bijection between non-negative
integer matrices and pairs of non-intersecting lattice paths.
Furthermore, the maximum displacement of the top-most of these paths
is equal to $L_n^{\symmU}$, with the profile of this path also
specifying the height profile in a statistical mechanical model
referred to as the polynuclear growth model. With regard to the latter, the
entries $x_{i,j}$ of an $n \times n$ non-negative integer matrix
$X = [x_{i,j}]_{i,j=1,\dots,n}$ (for convenience rows are labelled from
the bottom) now represent the heights of columns of unit width centred
about $x=j-i$ which occur at time $t=i+j-1$ (in labelling the matrix in terms
of $x$ and $t$ it is convenient to first rotate it 45${}^\circ$
anti-clockwise). The columns are to be placed on
top of the profile formed by earlier nucleation events and their growth.
The right boundary of the column corresponding to $x_{i,j}$ is to be
weighted $a_i^{x_{i,j}}$ while the left  boundary is to be weighted 
$b_j^{x_{i,j}}$, and these weights are to be multiplied together with
any existing weights along the same vertical segment of the profile.
During each time interval the existing profile or profiles are required
to grow one unit to the left and one unit to the right, with any resulting
overlap, together with the corresponding portion of the weights, recorded
on a profile with base one unit below. In this way a bijection between
$n \times n$ integer matrices with each entry $x_{i,j}$ weighted
$a_i b_j$, and a pair of weighted non-intersecting  paths is obtained.
A particular example is given in Figure \ref{bp.f15b}.

The bijection generates at most $n$ non-intersecting paths. These paths
start one unit apart in the $y$-direction (at $y=0,\dots,-(n-1)$). In the
$x$-direction one member of the pair of paths starts at $x=-(2n-3/2)$ and
may go up in arbitrary integer amounts of a unit step at
$x=-(2n+1/2-2j)$ for $j=1,\dots,n$ with each step weighted by $b_j$, or to
the right in steps of two units (these steps are unweighted). The other
member starts at $x=(2n-3/2)$ and may go up in arbitrary integer amounts
of a unit step at $x=(2n+1/2-2i)$ for $i=1,\dots,n$ with each step
weighted $a_i$, or to the left in steps of two units, the latter being
unweighted. Furthermore the second member is constrained so that it
joins with the first member at $x=0$. Both members are equivalent to what
will be termed u/rh (up/ right horizontal) non-intersecting lattice paths.
By definition such paths are defined on the square lattice and start at 
$x=0$, one unit apart in the $y$-direction (at $y=0,\dots,-(n-1))$, and
finish at $x=n-1$, with $y$-coordinates $\mu_l - (l-1)$
$(l=1,\dots,n)$ where $\mu_1 \ge \mu_2 \ge \cdots \ge \mu_N \ge 0$.
The path starting at $y=-(l-1)$ is referred to as the level-$l$ path.
Each path may move either up or to the right along the edges of the lattice,
with each up step at $x=j-1$ weighted $q_j$. Define the weight of a
configuration of u/rh lattice paths as the product of all the step weights.
Then it is well known (see e.g.~\cite{Sa01}) that with
$\mu := (\mu_1,\mu_2,\dots,\mu_N)$ (because of the orderings of the
$\mu_i$, $\mu$ forms a partition)
\begin{equation}\label{3.Sc}
\sum_{{\rm u/rh \, paths} \atop
{\rm displacements} \, \mu}
({\rm weight \: of \: the \: paths}) =
s_\mu(q_1,\dots,q_n)
\end{equation}
where $s_\mu$ is the Schur polynomial.

\begin{figure}
\epsfxsize=7cm
\centerline{\epsfbox{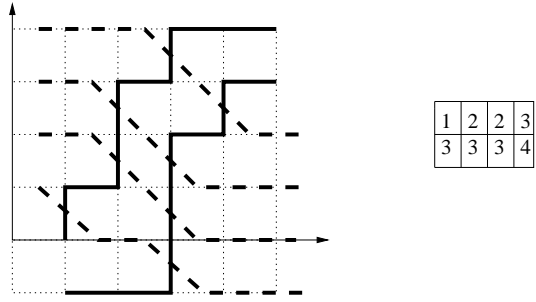}}
\caption{\label{bp.f15c} Drawn in heavy lines on the square lattice is a
family of two u/rh lattice paths with up segments allowed at $x=1,2,3,4$, 
while drawn in heavy dashed lines are the corresponding four dual
lattice paths. Also given is the semi-standard tableau encoding the
two u/rh lattice paths.
}
\end{figure}

We remark that u/rh non-intersecting lattice paths are
equivalent to semi-standard tableaux (numbered diagram of a partition
$\lambda$ such that the numbers weakly increase along rows and strictly decrease
down columns).
Thus with $\tilde{\lambda}_l(j)$ denoting the number of vertical steps in the
level-$l$ path at $x=j-1$, the $l$th row of the tableaux is of length
$\sum_{j=1}^n \tilde{\lambda}_l(j) =: \lambda_l$ and is numbered by
$\tilde{\lambda}_l(j)$ lots of $j$'s $(j=1,\dots,n$ in order). 
An explicit example is given in Figure \ref{bp.f15c}. Consequently if
there are $n$ lattice paths the numbering is from the set
$\{1,\dots,n\}$ which is referred to as the content of the tableau.
With $\lambda = (\lambda_1,\dots,\lambda_n)$ denoting the partition formed
from the length of the rows, the tableaux is said to have shape $\lambda$.

For future reference we note that with $\mu_l(n,j)$ denoting the displacement
of the level-$l$ path  at $x=-(2n+1/2-2j)$ 
as resulting from the RSK correspondence
and $\mu_l(i,n)$ equal to the
displacement of the level-$l$ path at $x=2n+1/2-2i$, we have
\begin{eqnarray}\label{4.1r}
\sum_{l=1}^n (\mu_l(n,j) - \mu_l(n,j-1) ) &=& \sum_{i=1}^n x_{i,j} \\
\sum_{l=1}^n (\mu_l(i,n) - \mu_l(i-1,n) ) &=& \sum_{j=1}^n x_{i,j}.
\end{eqnarray}
Also for future reference we make note of quantities generalizing
$L^\symmU_n$ which are related to the maximum displacements $\mu_l$ of the
level-$l$ paths for each $l=1,2,\dots$. For this let $({\rm rd}^*)^l$ denote
the set of $l$ disjoint (here meaning connecting no common lattice sites)
${\rm rd}^*$ lattice paths, the latter defined as either a single point,
or points connected by segments formed out of arbitrary positive integer
multiples of steps to the right and steps up in the square lattice
$1 \le i,j \le n$. Generalizing the definition (\ref{3.0}) by defining
\begin{equation}\label{4.2r}
L_n^{\symmU (l)} = {\rm max} \sum_{({\rm rd}^*)^l} x_{i,j},
\end{equation}
a theorem of Greene \cite{Gr74} gives
\begin{equation}\label{4.3r}
L_n^{\symmU(l)} = \sum_{m=1}^l \mu_m,
\end{equation}
and thus in particular \cite{Kn70}
\begin{equation}\label{2.10a}
\mu_1 = L_n^{\symmU (1)} := L_n^\symmU.
\end{equation}

It follows from the above discussion that with the entries of the 
$n \times n$ matrix $X$ weighted according to (\ref{2.1}), the probability
that $X$ maps under the  RSK correspondence to a pair of non-intersecting paths
with maximum displacements $\mu$ is given by \cite{Jo02}
\begin{equation}\label{6.1}
\prod_{i,j=1}^n(1 - a_i b_j) s_\mu(a_1,\dots,a_n)
s_\mu(b_1,\dots,b_n).
\end{equation}
The equality (\ref{2.10a}) between $\mu_1$ and $L_n^\symmU$ then implies 
the formula \cite{BR01a,Jo02}
\begin{equation}\label{6.2}
{\rm Pr}(L_n^\symmU \le l) = \prod_{i,j=1}^n(1 - a_i b_j)
\sum_{\mu:\mu_1 \le l}s_\mu(a_1,\dots,a_n)
s_\mu(b_1,\dots,b_n).
\end{equation}
Thus we must now show that the sum over Schur functions in (\ref{6.2}) can be
written as the average over $U(l)$ in (\ref{3.1}). Moreover we want to
achieve this task within the framework of non-intersecting paths.

An important notion for this purpose is that of the dual 
non-intersecting lattice paths associated with a set of u/rh paths.
The dual lattice paths connect points on the lattice
$\{(n-1/2,m), \, n \in \zz_{\ge 0}, m \in \zz \}$. Points are
connected by segments which are directed either right horizontal (rh)
or diagonally down (dd), with a dd segment bisecting every u segment
of the u/rh lattice path. The dd segments are connected by rh segments
to form dual lattice paths starting at $x=-1/2$ and finishing at $x=n-1/2$
in the $x$-direction, while in the $y$-direction these paths start at
$y=k$ for $k=1,2,\dots,\mu_1$ where $\mu_1$ is the maximum displacement
of the level-1 path (see Figure \ref{bp.f15c}). In terms of
tableaux, the dual lattice paths correspond to reading
down columns instead of across rows. The mapping carries over to
weighted paths by simply weighting the dd segment in the dual path
by the value of the u segment it bisects in the original u/rh path.

The key feature for our purpose is that the constraint on the u/rh
lattice paths having maximum displacement less than or equal to $l$
translates in the dual path picture as constraining the number of
paths to be less than or equal to $l$. As only dd segments are weighted
we can take the number of paths as being exactly equal to $l$.
Furthermore we can regard a pair of rh/dd  lattice paths, each with the
same end points and each containing $l$ paths but weighted from
$\{b_j\}$ and $\{a_i\}$ respectively, as a single set of 
non-intersecting paths. In the latter the second member of the pair
is reflected about $x=n-1/2$ so that its final positions are
at $(2n-1/2,k)$ for $k=1,2,\dots,l$ at it consists of right horizontal
and up diagonal segments, the latter being weighted by $a_j$ according 
to them passing through $x=2n+1/2-j$. With the initial and final
$y$ coordinates generalized to $y_1^{(0)},\dots, y_l^{(0)}$ and
$y_1,\dots,y_l$ respectively, let 
$G_{2n}(y_1^{(0)},\dots,y_l^{(0)};y_1,\dots,y_l)$ denote the total
weight of all such paths. It then follows from the dual lattice paths
interpretation of the sum in (\ref{6.2}) that
\begin{equation}\label{2.1r}
{\rm Pr}(L^\symmU \le l) = \prod_{i,j=1}^n(1-a_i b_j)
G_{2n}(\{y_j^{(0)}=-(j-1)\}_{j=1,\dots,l};
\{y_j=-(j-1)\}_{j=1,\dots,l}). 
\end{equation}
Because each path in the family is directed, the weight $G_{2n}$ of all
paths in the family is given in terms of the weight of a single path
in the family, $g_{2n}(y^{(0)};y)$ say. Thus the well known
Linstr\"om-Gessel-Viennot theorem (see e.g.~\cite{Kr95}) gives
\begin{equation}\label{3.1r}
G_{2n}
(y_1^{(0)},\dots,y_l^{(0)};y_1,\dots,y_l) = \det \Big [
g_{2n}(y^{(0)}_j;y_k)  \Big ]_{j,k=1,\dots,l}.
\end{equation}
Furthermore it is easy to see that
\begin{equation}\label{3.2r}
g_{2n}(y^{(0)};y) = {1 \over 2 \pi}
\int_0^{2 \pi} \prod_{j=1}^N(1 + a_j e^{-i\theta})(1+b_j e^{i\theta})
e^{-i(y-y^{(0)}) \theta} \, d\theta.
\end{equation}
Substituting (\ref{3.2r}) in (\ref{3.1r}), then substituting the result in
(\ref{2.1r}) and recalling the general formula \cite{Sz75}
\begin{equation}\label{2.14a}
\det \Big [ {1 \over 2 \pi} \int_0^{2 \pi} a(\theta)
e^{i(j-k) \theta} \, d \theta \Big ]_{j,k=1,\dots,n} =
\Big \langle \prod_{l=1}^n a(\theta_l) \Big \rangle_{U(n)}
\end{equation}
we see that (\ref{3.1}) is reclaimed.

\section{Polynuclear growth with Bernoulli random variables}
\setcounter{equation}{0}
A variant of the last passage percolation model revised in the previous section 
is to replace (\ref{2.1}) by the Bernoulli distribution
\begin{equation}\label{bp.ber}
{\rm Pr} (x_{ij} = k) = {(a_i b_j)^k \over 1 + a_i b_j}, \qquad k=0,1.
\end{equation}
Let $X=[x_{i,j}]_{i=1,\dots,m \atop j=1,\dots,n}$ be an array of such variables.
One specifies the corresponding last passage time by
\begin{equation}\label{bp.lm}
L_{m,n}^{01} := \max
\sum_{(i',j') \in {\rm bottom \, to \, top} \atop
{\rm u/rd \, paths}} x_{i',j'}
\end{equation}
where the sum is over all u/rd paths in the rectangle
$1 \le i' \le m$, $1 \le j' \le n$ from the bottom row
(row 1) to the top row (row $m$). The segments of the path
join entries successively to the north or north-east in the array.

Underlying this model is the dual RSK correspondence  \cite{Fu97}.
To our knowledge this has not previously been related to a layered growth
model formed out of non-intersecting paths. Here such a relationship will
be presented.

Again, the entries $x_{ij}$ of the array $X$ are regarded as recording
nucleation events. However, unlike the situation with the RSK correspondence
itself, the entries of $X$ are not first rotated 45${}^\circ$ before being
associated with positions and times. Rather
$x_{i,j} = 1$ denotes a nucleation event (a unit square) which is positioned
above the segment $x=j-1$ to $x=j$, and on top of earlier nucleation events
and their growth (this is in common with the polynuclear growth model of the
previous section). These nucleation events occur at successive times
$t=1,2,\dots,n$, with the positions recorded by 1's
in the corresponding rows of $X$. Thus to begin, at $t=1$ the nucleation events
are read off from the first row of $X$ and marked on the line $y=0$. As
$t \mapsto t+1$, the existing profile(s) is to grow one unit to the right
(but not the left) until it joins up with the neighbouring nucleation event
on the right. If there is no such right neighbour, and this nucleation 
event is yet to grow (i.e.~recorded in the previous time step), it
is to grow to $x=n+1$ and have its shape modified by removing the upper triangular
half of its final square. If it has right edge at $x \ge n+1$, it is to grow one
unit to the right. Also, the meeting of all nucleation events in going from
$t \mapsto t+1$ are to be recorded on the line $y=-t$ as new nucleation
events with left edge at the positions of the meetings.

\begin{figure}
\epsfxsize=12cm
\centerline{\epsfbox{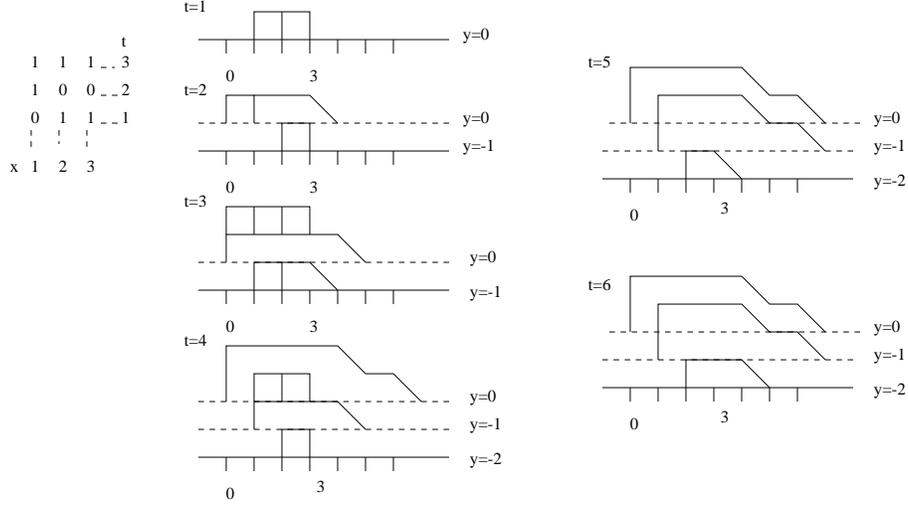}}
\caption{\label{bp.f6}
Mapping from a 0,1 matrix to a pair of
non-intersecting lattice paths. }
\end{figure}

This procedure is to stop after time $m+l$ along $y=-(l-1)$ $(l=1,\dots,m+1)$, this 
being the maximum time for which new nucleation events can be created and then grow
once. The layers of profiles which are so formed are of the form u/rh (up/ right
horizontal) non-intersecting paths from $x=0$ to $x=n-1$, and ld/lh (left diagonal/
left horizontal) non-intersecting paths from $x=n+m$ to $x=n+1$ (see Figure
\ref{bp.f6}).

\vspace{.5cm}

According to (\ref{3.Sc}) and the surrounding text the total weight of all
non-intersecting u/rh paths initially equally spaced at $y=0,\dots,-(n-1)$
along $x=0$, finishing at $y=\mu_j - (j-1)$, $(j=1,\dots,n)$ along $x=n-1$,
with up steps at $x=j-1$ weighted $b_j$ is given by $s_\mu(b_1,\dots,b_n)$.
Further, the dual paths of Figure \ref{bp.f15c} are precisely the non-intersecting
ld/lh paths initially equally spaced at $y=0,\dots,-(n-1)$ along $x=n+m$,
finishing at $y=\mu_j - (j-1)$ $(j=1,\dots,n)$ along $x=n+1$, which make up
the second family in the growth process. With the possible up steps (each of
unit length) at $x=n+m+1-i$ weighted by $a_i$, the
total weight of the paths is $s_{\mu'}(a_1,\dots,a_m)$, where $\mu'$ denotes the
partition corresponding to the transpose of the diagram of $\mu$.

It follows from this that with an $n \times m$ array of $0$'s and $1$'s formed
according to (\ref{bp.ber}), the probability that it maps, under the dual
RSK correspondence, to the above specified nested growth profile with
maximum displacement $\mu$ is given by
\begin{equation}\label{bp.du}
\prod_{i=1}^m \prod_{j=1}^n (1 + a_i b_j)^{-1}
s_{\mu'}(a_1,\dots,a_m) s_\mu(b_1,\dots,b_n)
\end{equation}
(cf.~(\ref{6.1})). Analogous to (\ref{6.2}) we have the
normalization condition
\begin{equation}\label{bp.du1}
\prod_{i=1}^m \prod_{j=1}^n (1 + a_i b_j)^{-1}
\sum_{\mu} s_{\mu'}(a_1,\dots,a_m) s_\mu(b_1,\dots,b_n) = 1.
\end{equation}
It follows immediately from (\ref{bp.du}) that
\begin{equation}\label{bp.du2}
{\rm Pr}(L_{m,n}^{01} \le l) =
\prod_{i=1}^m \prod_{j=1}^n (1 + a_i b_j)^{-1}
\sum_{\mu_1 \le l} s_{\mu'}(a_1,\dots,a_m)
s_\mu(b_1,\dots,b_n).
\end{equation}

From Figure \ref{bp.f15c} and related text, we know that the Schur polynomial
$s_{\mu'}$ can be interpreted in terms of non-intersecting u/rh paths by reading
along rows, while $s_{\mu}$ can be interpreted in terms of u/rd paths by reading
down columns. In both cases the number of paths is equal to $\mu_1$, which according
to (\ref{bp.du2}) is no bigger than $l$. Further, the left set of paths consists of
$m$ steps, and the right set consists of $n$ steps. Let $G_{n,m}^*(\vec{l}^{(0)};
\vec{l}^{(0)})$, $\vec{l}^{(0)} := (l-1,l-2,\dots,0)$ denote the total weight of such
paths. Analogous to (\ref{2.1r}) we have that
\begin{equation}\label{2.ex}
{\rm Pr}(L_{m,n}^{01} \le l) =
\prod_{i=1}^m \prod_{j=1}^n (1 + a_i b_j)^{-1} G^*_{n,m}(\vec{l}^{(0)};
\vec{l}^{(0)}).
\end{equation}
But according to the Linstr\"om-Gessel-Viennot theorem
\begin{equation}\label{2.13'}
G^*_{n,m}(\vec{l}^{(0)};
\vec{l}^{(0)}) = \det [ g_{n,m}^*({l}_j^{(0)};{l}_k^{(0)}) ]_{j,k=1,\dots,l}
\end{equation}
where $ g_{n,m}^*(x,y)$ is the weight of a single path of the prescribed type starting
at $x$ and finishing at $y$. The latter can readily be seen to be given by
\begin{equation}\label{2.14'}
 g_{n,m}^*(x;y) = {1 \over 2 \pi}
\int_0^\pi \prod_{j=1}^m (1 + a_j e^{i \theta})
\prod_{k=1}^n(1 - b_k e^{- i \theta})^{-1} e^{- i \theta (y-x) } \, d \theta
\end{equation}
Substituting (\ref{2.14'}) in (\ref{2.13'}), making use of (\ref{2.14a}), and 
substituting in (\ref{2.ex}) we reclaim the expression for ${\rm Pr}(L_{m,n}^{01} \le l)$
as a random matrix average \cite{BR01a},
\begin{equation}
{\rm Pr}(L_{m,n}^{01} \le l) 
 =
\prod_{i=1}^m \prod_{j=1}^n (1 + a_i b_j)^{-1}
\Big \langle \Big (
\prod_{j=1}^m \prod_{k=1}^l(1 + a_j e^{i \theta_k}) \Big )
 \Big (
\prod_{j=1}^n \prod_{k=1}^l(1 - b_j e^{-i \theta_k}) \Big )^{-1}
\Big
\rangle_{{\rm CUE}_l}.
\end{equation}

\section{Matrices symmetric about the anti-diagonal}
\setcounter{equation}{0}
With our convention of numbering rows from the bottom, the term 
anti-diagonal used here is what is conventionally referred to as the
diagonal of the matrix. Under the RSK correspondence in the non-intersecting paths
formulation, matrices symmetric about the anti-diagonal give a bijection
with  pairs of non-intersecting u/rd lattice paths in which only one
member of the pair is independent.

Regarding this point, reflect the entries of a general $n \times n$
non-negative integer $X$ about the anti-diagonal to form the matrix
$X^R = [x_{n+1-j,n+1-i}]_{i,j=1,\dots,n}$. We see from the definition
(\ref{4.2r}) that the quantities $L_n^{\symmU(l)}$ are invariant under this
transformation, and thus according to (\ref{4.3r}) so then too are the
path displacements $\mu_l$. Furthermore, it follows from (\ref{4.1r}) that
\begin{eqnarray}\label{5.sn}
&& \sum_{l=1}^n ( \mu_l^R(n,j) - \mu_l^R(n,j-1) ) =
\sum_{i=1}^n x_{n+1-j,i} = \sum_{l=1}^n ( \mu_l(n+1-j,n) -
\mu_l(n-j,n) ) \nonumber \\
&& \sum_{l=1}^n ( \mu_l^R(i,n) - \mu_l^R(i-1,n) ) =
\sum_{i=1}^n x_{j,n+1-i} = \sum_{l=1}^n ( \mu_l(n,n+1-j) -
\mu_l(n,n-j) ).
\end{eqnarray}
These equations respectively tell us that the total number of up steps at
$x=-2n-{1 \over 2} +2j$ $(x=2n+{1 \over 2} -2i)$ in the paths corresponding to
$X^R$ is equal to the number of up steps at $x=2j-{3 \over 2}$
($x=-2i+{3 \over 2}$) in the paths corresponding to $X$. Furthermore, if
$X=X^R$ an algorithm can be presented (due to Sch\"utzenburger; see
e.g.~\cite{Sa01}) which allows paths from $x=2n-1/2$ to $x=1/2$ to 
be constructed out of the paths from $x=-2n +1/2$ to $x=-1/2$. This
permits a bijection between weighted matrices symmetric about the
diagonal and a single set of u/rh lattice paths provided the weighting
of $x_{i,j}$ in the former is equal to $(a_i a_{n+1-j})^{x_{i,j}}$
(and thus $b_{n+1-i} = a_i$). For example, the matrix of 
Figure \ref{bp.f15b} is symmetric about the anti-diagonal. With $n=2$,
setting
$b_{n+1-i} = a_i$ $(i=1,\dots,n)$ we see that the weight of steps at
$x=-2n-1/2+2i$ is equal to the weight of up steps at $x=2i-1/2$
($i=1,\dots,n$).

With the constraint $X = X^R$, to obtain a probabilistic setting we should
weight only the sites $i \le n+1-j$. To achieve this without affecting the
weights of the pairs of paths we simply square the weights at the sites 
$i < n+1-j$, and set the weights at sites $i > n+1-j$ to unity. With
$a_i = \sqrt{q_i}$, we therefore choose
\begin{eqnarray}\label{6.1r} 
&& {\rm Pr}(x_{i,j} = k) = (1 - q_i q_{n+1-j}) (q_i q_{n+1-j})^k,
\qquad i < n+1-j, \nonumber \\&&
{\rm Pr} (x_{i,n+1-i} = k) = (1 - q_i) q_i^k
\end{eqnarray}
which corresponds to weighting the vertical segments of the single u/rh
paths by $q_1,q_2,\dots, q_n$ from right to left. According to 
(\ref{3.Sc}) with the maximum displacement of the level-$l$ path
denoted by $\mu_l$, the total weight of such paths is given by
$s_\mu(q_1,\dots,q_n)$. We conclude that under the RSK mapping, with the
independent entries chosen according to (\ref{6.1r}), the probability a
non-negative integer matrix symmetric about the anti-diagonal maps to a
pair of u/rh paths with final displacement $\mu$ is equal to
\begin{equation}\label{6.2r}
\prod_{i=1}^n(1 - q_i) \prod_{1 \le i < j \le n} (1 - q_i q_j)
s_\mu(q_1,\dots,q_n).
\end{equation}

The specification (\ref{6.1r}) can be generalized, allowing for a
generalization of (\ref{6.2r}). For this one recalls \cite{Kn70}
that the RSK correspondence has the property that for $[x_{i,j}]$
symmetric about the anti-diagonal
\begin{equation}\label{6.2r'}
\# \{ x_{i,n+1-i}: \, x_{i,n+1-i} \: {\rm odd} \} = \# \{ \mu_j:
\, \mu_j \: {\rm odd} \} =
\sum_{j=1}^n (-1)^{j-1} \mu_j'
\end{equation}
where $\mu_j'$ denotes the displacement of the level $j$ conjugate
path or equivalently the length of the $j$th column in the
corresponding tableau (recall Figure \ref{bp.f15c}). 
Hence if we generalize the second
probability in (\ref{6.1r}) to read
\begin{equation}\label{6.7r}
{\rm Pr} (x_{i,n+1-i} = k) = {(1 - q_i^2) \over 1 + \beta q_i}
\beta^{k \, {\rm mod} \,2} q_i^k
\end{equation}
then we have that (\ref{6.2r}) generalizes to
$$
\prod_{i=1}^n{(1 - q_i^2) \over 1 + \beta q_i}
\prod_{1 \le i < j \le n} (1 - q_i q_j)
\beta^{\sum_{j=1}^n(-1)^{j-1} \mu_j'}
s_\mu(q_1,\dots,q_n).
$$
Writing
$$
L^{\symmS}_n := {\rm max} \sum_{(1,1) {\rm u/rh} (n,n) \atop X = X^R} x_{i,j}
$$
it follows from this that \cite{BR01a}
\begin{equation}\label{6.71}
{\rm Pr}(L^{\symmS}_n \le l) = \prod_{i=1}^n {1 - q_i^2 \over 1 + \beta q_i}
\prod_{1 \le i < j \le n} (1 - q_i q_j)
\sum_{\mu: \mu_1 \le l} \beta^{\sum_{j=1}^n (-1)^{j-1} \mu_j'}
s_\mu(q_1,\dots,q_n).
\end{equation}
Using symmetric function theory, Baik and Rains \cite{BR01a} show that the
sum in (\ref{6.71}) can be written as a random matrix average involving the
classical group $Sp(2l)$. Matrices from this subgroup of $U(2l)$ have their
eigenvalues in complex conjugate pairs $\{e^{\pm i \theta_j} \}_{j=1,\dots,l}$,
with $0 < \theta_j < \pi$ $(j=1,\dots,l)$. Here we will give a derivation
in keeping with integration techniques from random matrix theory.

\begin{prop}\label{p7a}
Consider the eigenvalues with angles $0 < \theta_j < \pi$ $(j=1,\dots,l)$ of
matrices from Sp$(2l)$. 
Let $\langle \: \rangle_{Sp(2l)}$ denote an average with respect to the
corresponding 
eigenvalue p.d.f.,
\begin{equation}\label{6.Sp}
{1 \over (2 \pi)^l} {1 \over 2^l l!}
\prod_{k=1}^l |e^{i \theta_k} - e^{-i \theta_k}|^2
\prod_{1 \le j < k \le l} |e^{i \theta_j} - e^{i \theta_k} |^2
|1 - e^{i (\theta_j + \theta_k)} |^2.
\end{equation}
One has
\begin{eqnarray}
&& {\rm Pr}(L^{\symmS}_n \le 2l) =
\prod_{i=1}^n {1 - q_i^2 \over 1 + \beta q_i} \prod_{1 \le i < j \le n}
(1 - q_i q_j) \Big \langle \prod_{k=1}^l \Big (
{1 \over | 1 - \beta e^{-i \theta_k}|^2} \prod_{j=1}^n
|1 + q_j e^{i \theta_k}|^2  \Big ) \Big \rangle_{Sp(2l)}
\label{8.89} \\
&& {\rm Pr}(L^{\symmS}_n \le 2l+1) =
\prod_{i=1}^n (1 - q_i^2)  \prod_{1 \le i < j \le n}
(1 - q_i q_{n+1-j})
\Big \langle \prod_{k=1}^l \prod_{j=1}^n |1 + q_j e^{i \theta_k}|^2
\Big \rangle_{Sp(2l)}. \label{8.90}
\end{eqnarray}
\end{prop}

\noindent
{\bf Proof} \quad The maximum possible height in the growth model relating to
(\ref{bp.du2}) is $m$, implying the so called dual Cauchy identity
\begin{equation}\label{dC}
\prod_{i=1}^m \prod_{j=1}^n (1 + a_i b_j) =
\sum_{\mu_1 \le m} s_{\mu'}(a_1,\dots,a_m)
s_\mu(b_1,\dots,b_n).
\end{equation}
Renaming the parameters, it follows from this that
\begin{equation}\label{bp.apr}
\prod_{k=1}^l \prod_{j=1}^n |1 + q_j e^{i \theta_k}|^2 =
\sum_{\mu: \mu_1 \le 2l}
s_\mu(q_1,\dots,q_n) s_{\mu'}(e^{i \theta_1}, e^{-i \theta_1}, \dots, e^{i \theta_l}, e^{-i \theta_l}).
\end{equation}
Substituting (\ref{bp.apr}) in (\ref{8.89}), substituting the result in (\ref{6.71}), equating
coefficients of $s_\mu(q_1,\dots,q_n)$ and writing $\mu'=\rho$ shows that (\ref{8.89}) is
equivalent to the matrix integral formula
\begin{equation}\label{bp.Sp}
\Big \langle \prod_{k=1}^l {1 \over |1 - \beta e^{-i \theta_k}|^2}
s_{\rho}(e^{i \theta_1},e^{-i \theta_1},\dots,e^{i \theta_l},e^{-i \theta_l})
\Big \rangle_{ Sp(2l)} = \beta^{\sum_{j=1}^{2l} (-1)^{j-1} \rho_j}.
\end{equation}
Now, the meaning of the matrix integral is an integral over $\theta_k \in [0,\pi]$, weighted by
(\ref{6.Sp}). Noting that the integrand is unchanged by $\theta_l \mapsto - \theta_l$,
and making use of the determinant formula for Schur polynomials
\begin{equation}\label{sdet}
s_\lambda(x_1,\dots,x_N) = {\det [ q_j^{N - k + \lambda_k} ]_{j,k=1,\dots,N} \over
\det [ q_j^{N-k} ]_{j,k=1,\dots,N} }.
\end{equation}
shows that the matrix integral is equal to
\begin{equation}\label{bp.Sp1}
{1 \over (2 \pi)^l 2^l l!} \int_{-\pi}^\pi d \theta_1 \cdots \int_{-\pi}^\pi d \theta_l \,
\prod_{k=1}^l {(e^{i \theta_k} - e^{-i \theta_k} ) \over |1 - \beta e^{-i \theta_k} |^2 }
\det \left [ \begin{array}{l} e^{i \theta_j(\rho_{2l-k+1} + k -1)} \\
 e^{- i \theta_j(\rho_{2l-k+1} + k -1)} \end{array} \right ]_{j=1,\dots,l \atop k=1,\dots,2l}.
\end{equation}

The structure of the integral (\ref{bp.Sp1}) is very common in random matrix theory
\cite{Me91,Fo02}. It can be written as a Pfaffian, giving that (\ref{bp.Sp1}) is equal to
\begin{equation}\label{bp.Sp2}
{1 \over 2^l} {\rm Pf} \Big [ {1 \over 2 \pi} \int_{-\pi}^\pi
{e^{i \theta} - e^{-i \theta} \over |1 - \beta e^{- i \theta} |^2}
(e^{i \theta (\rho_j - j - \rho_k + k)} - e^{- i \theta (\rho_j - j - \rho_k + k)} )
d \theta
\Big ]_{j,k=1,\dots,2l},
\end{equation}
which after evaluating the integral therein reduces to
\begin{equation}\label{bp.Sp7}
\beta^{-l} {\rm Pf} \Big [ {\rm sgn}(k-j) \beta^{|\rho_j - \rho_k + k - j|}
\Big ]_{j,k=1,\dots,2l}.
\end{equation}
This Pfaffian is special case $x_j = \rho_j - j$, $f(x_j) = \beta^{x_j}$ in the general formula
\cite{FR02}
\begin{equation}\label{bp.Sp3}
 {\rm Pf} \Big [ \Big ( {f(x_j) \over f(x_k)} \Big )^{{\rm sgn}(x_j - x_k)}
{\rm sgn}(x_j - x_k) \Big ]_{j,k=1,\dots,2l} =
\prod_{j=1}^l {f(x_{Q(2j-1)}) \over f(x_{Q(2j)}) } \varepsilon(Q),
\end{equation}
where the permutation $Q$ is such that
$$
x_{Q(2j-1)} > x_{Q(2j)}, \qquad
Q(2j) > Q(2j-1) \: \: (j=1,\dots,l)
$$
and thus evaluating to the
r.h.s.~of (\ref{bp.Sp}).

In regards to (\ref{8.90}), use  of an appropriate modification of (\ref{bp.apr}) shows
that this is equivalent to the matrix integral formula
\begin{equation}\label{bp.Spe}
\Big \langle
s_{\rho}(e^{i \theta_1},e^{-i \theta_1},\dots,e^{i \theta_l},e^{-i \theta_l},\beta)
\Big \rangle_{ Sp(2l)} = \beta^{\sum_{j=1}^{2l+1} (-1)^{j-1} \rho_j}.
\end{equation}
To derive this,
use of (\ref{6.Sp}) and (\ref{sdet}) shows that the matrix integral is equal to
\begin{eqnarray*}
&&
{1 \over (2 \pi)^l 2^l l!} \int_{-\pi}^\pi d \theta_1 \cdots \int_{-\pi}^\pi d \theta_l \,
\prod_{k=1}^l {(e^{i \theta_k} - e^{-i \theta_k} ) \over |1 - \beta e^{-i \theta_k} |^2 }
\det \left [ \begin{array}{l} e^{i \theta_j(\rho_{2l-k+2} + k -1)} \\
 e^{- i \theta_j(\rho_{2l-k+2} + k -1)} \\
\beta^{\rho_{2l-k+2} + k -1}
\end{array} \right ]_{j=1,\dots,l \atop k=1,\dots,2l+1}
\\
&& = \quad
{1 \over 2^l} {\rm Pf} \left[
\begin{array}{cc}
A_{(2l+1) \times (2l+1)}
& [\beta^{\rho_j+2l+1-j}]_{j=1,\dots,2l+1} \\
{}[-\beta^{\rho_k+2l+1-k}]_{k=1,\dots,2l+1} & 0 \end{array} \right ].
\end{eqnarray*}
where
$$
A_{(2l+1) \times (2l+1)} := \left [ {1 \over 2 \pi} \int_{-\pi}^\pi
{e^{i \theta} - e^{-i \theta} \over |1 - \beta e^{- i \theta} |^2}
(e^{i \theta (\rho_j - j - \rho_k + k)} - e^{- i \theta (\rho_j - j - \rho_k + k)} )
d \theta
\right ]_{j,k=1,\dots,2l+1}.
$$
Here the second equality follows from standard integration methods of random matrix
theory, in the same way that (\ref{bp.Sp2}) follows from  (\ref{bp.Sp1}).
Computing the integral reduces this to
$$
\beta^{-(l+1)} {\rm Pf} \left [
\begin{array}{cc}
[{\rm sgn}(k-j) \beta^{|\rho_j - \rho_k + k - j|}
 ]_{j,k=1,\dots,2l+1} &
[ \beta^{\rho_j + 2l + 1 - j} ]_{j=1,\dots,2l} \\
{}[- \beta^{\rho_k + 2l + 1 - k} ]_{k=1,\dots,2l} & 0
\end{array}
\right ].
$$
But this is precisely the same as (\ref{bp.Sp7}) with $l \mapsto l + 1$, $\rho_{2l+2}=0$,
and so reduces to the r.h.s.~of (\ref{bp.Spe}).
\hfill $\square$

\section{Matrices symmetric about the diagonal}
\setcounter{equation}{0}
According to the rules of the polynuclear growth model, if a
non-negative integer matrix $X=[x_{i,j}]_{i,j=1,\dots,n}$ maps to a
pair of u/rh non-intersecting lattice paths $(P_1,P_2)$
of the same final displacement, then the
transposed matrix $X^T= [x_{j,i}]_{i,j=1,\dots,n}$ maps to the pair of
u/rh non-intersecting paths $(P_2,P_1)$. Hence the
Robinson-Schensted-Knuth correspondence when applied to symmetric matrices
$X=X^T$ gives a bijection with a single set of u/rh lattice paths,
since then we must have $P_1=P_2$. To obtain a bijection between weighted
symmetric matrices and a weighted set of u/rh lattice paths, and furthermore
to obtain a probabilistic setting, we weight only the entries $i \le j$, with
the value of $x_{i,j}$ for $i > j$ fixed by symmetry. Arguing then as in the
derivation of (\ref{6.2r}) we see with
\begin{equation}\label{8.103}
{\rm Pr}(x_{i,j}=k) = (1 - q_iq_j)(q_iq_j)^k, \: \: i < j
\qquad {\rm Pr}(x_{i,i}=k) = (1 - q_i) q_i^k
\end{equation}
the probability the symmetric  matrix $X$
maps to a set of u/rh paths with final displacements $\mu$ is equal to
\cite{Jo99a}
\begin{equation}\label{8.104}
\prod_{i=1}^n(1 - q_i)
\prod_{1 \le i < j \le n} (1 - q_i q_j)
s_\mu(q_1,\dots,q_n).
\end{equation}

As with (\ref{6.2r}) this can be generalized to the case that the
diagonal entries are chosen according to
\begin{equation}\label{8.106}
{\rm Pr}(x_{i,i}=k) = (1 - \alpha q_i) (\alpha q_i)^k.
\end{equation}
Thus recalling \cite{Kn70,FR02b} that in the Robinson-Schensted-Knuth
correspondence for symmetric matrices
\begin{equation}\label{bp.jog2}
\sum_{j=1}^n x_{j,j} = \sum_{j=1}^n (-1)^{j-1} \mu_j,
\end{equation}
with the generalization (\ref{8.106}), (\ref{8.104}) should correspondingly be
generalized to read \cite{BR01a}
\begin{equation}\label{8.107}
\prod_{i=1}^n(1 -  \alpha q_i) \prod_{1 \le i < j \le n} (1 - q_i q_j)
\alpha^{\sum_{j=1}^n(-1)^{j-1} \mu_j}
s_\mu(q_1,\dots,q_n).
\end{equation}
Writing
$$
L^{\symmO}_n := {\rm max} \sum_{(1,1) {\rm u/rh} (n,n) \atop X=X^T} x_{i,j}
$$
and noting that
$$
\sum_{j=1}^n (-1)^{j-1} \mu_j = \# ({\rm columns \: of \: odd \:
length \: in} \: \mu) =
\sum_{k=1}^l \mu_k' {\rm mod} \,2
$$
where $l=\mu_1$, it follows from (\ref{8.107}) that
\begin{equation}\label{8.109}
{\rm Pr}(L^\symmO_n \le l) =
\prod_{i=1}^n(1 -  \alpha q_i) \prod_{1 \le i < j \le n} (1 - q_i q_j)
\sum_{\mu: \mu_1 \le l}
\alpha^{\sum_{k=1}^l \mu_k' {\rm mod} \,2}
s_\mu(q_1,\dots,q_n).
\end{equation}
Starting with this formula, Baik and Rains \cite{BR01a} proved the following
analogue of Proposition \ref{p7a}, involving now a random matrix average 
involving the classical group $O(l)$. Matrices from this subgroup of $U(l)$
form two disjoint components, $O^+(l)$ and $O^-(l)$, distinguished by the
determinant equalling $+1$ or $-1$ respectively.
The complex eigenvalues occur in complex conjugate pairs, and there is a real
eigenvalue $z=-1$ for matrices in $O^-(l)$ with $l$ odd, a real eigenvalue
eigenvalue $z=1$ for matrices in $O^+(l)$ with $l$ even, and two real 
eigenvalues
$z=\pm 1$ for matrices in $O^-(l)$ with $l$ even.

\begin{prop}\label{p8a}
Consider the eigenvalues with angles $0 < \theta_j < \pi$, $(j=1,\dots,l)$ of matrices 
from $O(l)$.
Define
$$
\langle \: \cdot \: \rangle_{O(l)} = {1 \over 2} \Big (
\langle \: \cdot \: \rangle_{O^+(l)} +
\langle \: \cdot \: \rangle_{O^-(l)} \Big )
$$
where $\langle \: \cdot \: \rangle_{O^+(l)}$ denotes an average with respect
to the eigenvalue p.d.f.~for random matrices from the classical group
$O^+(l)$,
\begin{eqnarray}
&& {1 \over \pi^{l/2} 2^{l-1} (l/2)!}
\prod_{1 \le j < k \le l/2} |e^{i \theta_j} - e^{i \theta_k}|^2
|1 - e^{i(\theta_j + \theta_k)}|^2, \qquad l \: {\rm even} \label{O.1}\\
&& {1 \over \pi^{(l-1)/2} 2^{l-1} ((l-1)/2)!}
  \prod_{j=1}^{(l-1)/2} |1- e^{i \theta_j}|^2
\prod_{1 \le j < k \le (l-1)/2}|e^{i \theta_j} - e^{i \theta_k}|^2
|1 - e^{i(\theta_j + \theta_k)}|^2, \qquad l \: {\rm odd},
\nonumber \\
\end{eqnarray}
and $\langle \: \cdot \: \rangle_{O^-(l)}$ denotes an average with respect
to the eigenvalue p.d.f.~for random matrices from the classical group
$O^-(l)$,
\begin{eqnarray}
&& {1 \over \pi^{l/2-1} 2^{l-2}  (l/2)!} 
 \prod_{k=1}^{l/2 - 1} |1 - e^{2i \theta_k}|^2
\prod_{1 \le j < k \le l/2-1} |e^{i \theta_j} - e^{i \theta_k}|^2
|1 - e^{i(\theta_j + \theta_k)}|^2, \qquad l \: {\rm even} \nonumber \\
\label{O.2}\\
&& {1 \over \pi^{(l-1)/2} 2^{l-1} ((l-1)/2)!}
\delta(\theta_l - \pi) \prod_{j=1}^{(l-1)/2} |1+ e^{i \theta_j}|^2
\prod_{1 \le j < k \le (l-1)/2}|e^{i \theta_j} - e^{i \theta_k}|^2
|1 - e^{i(\theta_j + \theta_k)}|^2, \qquad l \: {\rm odd}.
\nonumber \\
\end{eqnarray}
We have
\begin{equation}\label{8.110}
{\rm Pr}(L^\symmO_n \le l) =
\prod_{i=1}^n(1 -  \alpha q_i) \prod_{1 \le i < j \le n} (1 - q_i q_j)
\Big \langle \det({\bf 1}_l + \alpha  U) \prod_{j=1}^n({\bf 1}_l + q_j
 U) 
\Big \rangle_{ U \in {\rm O}(l)}.
\end{equation}
\end{prop}

\noindent
{\bf Proof.} \quad Use of the dual Cauchy identity (\ref{dC})
in (\ref{8.110}) and comparison with (\ref{8.109}) shows
that (\ref{8.110}) is equivalent to the matrix integral evaluation
\begin{equation}\label{bp.OU}
\Big \langle \det ( {\bf 1}_l + \alpha  U)
s_\rho( U) \Big \rangle_{{ U} \in O(l)} = \alpha^{\sum_{j=1}^l \rho_j {\rm mod} \,2},
\end{equation}
where $s_\rho( U)$ denotes the Schur polynomial as a function of all the eigenvalues
of $U$.

Consider first the $l$ even case, $l \mapsto 2l$, and consider separately
the components $O^\pm(2l)$ of $O(2l)$. Substituting the eigenvalue p.d.f.~for
$O^+(2l)$ (\ref{O.1}), and proceeding as in the derivation of (\ref{bp.Sp7}),
which involves use of (\ref{sdet}) and reduction to a Pfaffian, shows
\begin{equation}\label{bp.106a}
\langle \det ({\bf 1}_{2l} + \alpha  U) s_\rho( U) \rangle_{{ U} \in O^+(2l)} =
2^{1 - l} {\rm Pf} [ a_{jk} ]_{j,k=1,\dots,2l}
\end{equation}
where
\begin{eqnarray}
a_{jk} & = & \Big ( (1 + \alpha^2) \delta_{(\rho_j - j) - (\rho_k - k), {\rm odd}} +
2 \alpha  \delta_{(\rho_j - j) - (\rho_k - k), {\rm even}} \Big )
{\rm sgn} (k-j) \nonumber \\ & = &
\Big ( {1 \over 2} (1 + \alpha)^2 -  {1 \over 2} (1 - \alpha)^2
(-1)^{(\rho_j - j) - (\rho_k - k)} \Big ) {\rm sgn} (k-j). \label{5.13a}
\end{eqnarray}
The task is therefore to compute the Pfaffian of the matrix with these entries. For
this one uses the identity \cite{St90}
\begin{equation}\label{bp.107}
{\rm Pf}( A +  B) =
\sum_{S \subseteq \{1,2,\dots,2l\} \atop |S| \: \: {\rm even} }
(-1)^{\sum_{j \in S} j - |S|/2} {\rm Pf}_S ( A) {\rm Pf}_{\bar{S}}( B)
\end{equation}
where $ {\rm Pf}_S ( A)$ denotes the Pfaffian of $ A$ restricted to rows and
columns specified by the index set $S$, and similarly ${\rm Pf}_{\bar{S}}( B)$.
With
$$
 A = \Big [ {1 \over 2} (1 + \alpha)^2 {\rm sgn} (k-j) \Big ]_{j,k=1,\dots,2l},
\qquad
 B = \Big [ - {1 \over 2} (1 - \alpha)^2
(-1)^{(\rho_j - j) - (\rho_k - k)}
{\rm sgn} (k-j) \Big ]_{j,k=1,\dots,2l},
$$
and noting
$$
{\rm Pf}[ {\rm sgn} (k-j) ] = 1, \qquad
{\rm Pf}[ a_{j,k} (-1)^{(\rho_j - j) - (\rho_k - k)} ] =
(-1)^{\sum (\rho_j - j) } {\rm Pf} [a_{j,k}]
$$
we see that
$$
{\rm Pf}_S  A = 2^{-|S|/2} (1 + \alpha)^{|S|}, \qquad
{\rm Pf}_{\bar{S}}  B = (-2)^{-|\bar{S}|/2} (1 - \alpha)^{2l - |S|}
(-1)^{\sum_{j \in \bar{S}} \rho_j - j}.
$$

It thus follows from (\ref{bp.107}) that 
\begin{equation}\label{bp.108}
2^{1 - l} {\rm Pf} ( A +  B) =
2 \sum_{S \subseteq \{1,2,\dots,2l\} \atop |S| \: \: {\rm even} }
\Big ( {1 + \alpha \over 2} \Big )^{|S|}
\Big ( {1 - \alpha \over 2} \Big )^{2l - |S|}
(-1)^{\sum_{j \in \bar{S} } \rho_j }.
\end{equation}
Now, in general
\begin{eqnarray*}
&& \sum_{S \subseteq \{1,2,\dots,2l\} \atop |S| \: \: {\rm even} }
x^{|S|} y^{2l - |S|} (-1)^{\sum_{j \in \bar{S}} \rho_j}
\nonumber \\
&& \qquad
 =
{1 \over 2} \Big (
\sum_{S \subseteq \{1,2,\dots,2l\} }
x^{|S|} y^{2l - |S|} (-1)^{\sum_{j \in \bar{S}} \rho_j} +
\sum_{S \subseteq \{1,2,\dots,2l\} }
x^{|S|} (-y)^{2l - |S|} (-1)^{\sum_{j \in \bar{S}} \rho_j} \Big ) \\
 && \qquad  =
{1 \over 2} \Big (
\prod_{j=1}^{2l} (x + (-1)^{\rho_j} y ) + \prod_{j=1}^{2l} (x - (-1)^{\rho_j} y ) \Big ).
\end{eqnarray*}
Using this result to evaluate (\ref{bp.108}) and substituting in (\ref{bp.106a}) gives the
matrix integral evaluation
\begin{equation}\label{bp.109}
\langle \det ( {\bf 1}_{2l} + \alpha  U ) s_\rho( U) \rangle_{{ U} \in O^+(2l)} =
\alpha^{\sum_{j=1}^{2l} \rho_j {\rm mod} \, 2} + \alpha^{\sum_{j=1}^{2l}
(\rho_j + 1) {\rm mod} \, 2}.
\end{equation}

We turn now to the corresponding formula for the average over $O^-(2l)$.
The analogue of (\ref{bp.106a}) in this case is
\begin{equation}\label{8.108a}
\langle \det ({\bf 1}_{2l} + \alpha  U) s_\rho( U) \rangle_{U \in O^-(2l)}
 = {(1 - \alpha^2) \over 2^{l-1}} [ \zeta]
{\rm Pf}[a_{j,k} + \zeta b_{j,k} ]_{j,k=1,\dots,2l}
\end{equation}
where $a_{j,k}$ is as in (\ref{bp.106a}) while
$b_{j,k} = (-1)^{\rho_k - k} - (-1)^{\rho_j - j}$, and $[\zeta]$ denotes the coefficient
of $\zeta$.
 Observing that
\begin{equation}\label{bp.108b}
[b_{jk}] = \vec{u} \vec{w}^T - \vec{w} \vec{u}^T, \qquad
\vec{u} = [1]_{j=1,\dots,2l}, \: \: \vec{w} = [(-1)^{\rho_j - j}]_{j=1,\dots,2l}
\end{equation}
shows that $[b_{jk}]$ has rank 2. It follows  that the Pfaffian in
(\ref{8.108a}) is linear in $\zeta$, and so the r.h.s.~of (\ref{8.108a}) can be rewritten
\begin{equation}\label{bp.108c}
{(1 - \alpha^2) \over 2^{l-1} } {1 \over \zeta} \Big (
{\rm Pf} \Big [ [a_{jk} ] + \zeta [b_{jk}] \Big ] - {\rm Pf} [a_{jk}] \Big ).
\end{equation}
With $\gamma, \zeta_1,\zeta_2$ arbitrary non-zero constants,
the structure of $[b_{jk}]$  and use of elementary row and column operations
verifies that this in turn can be rewritten
\begin{equation}\label{bp.109e}
{(1 - \alpha^2) \over 2^{l-1} } {1 \over \zeta_1 \zeta_2} \bigg (
{\rm Pf} \left [ \begin{array}{ccc} [a_{jk} ] & \zeta_1 \vec{w} & \zeta_2 \vec{u} \\
- \zeta_1 \vec{w}^T & 0 & \gamma \\
- \zeta_2 \vec{u}^T & - \gamma & 0 \end{array} \right ] -
\gamma {\rm Pf} [a_{jk}] \bigg ).
\end{equation}
Setting
$
\zeta_1 = {1 \over 2} (1 - \alpha)^2, \qquad \zeta_2 = (1 + \alpha)^2,
$
adding one half of the final row/column to the second last row/column, and subtracting
the second last row/column from the final row column, then setting $\gamma = (1 + \alpha^2)$
allows (\ref{bp.109e}) to be rewritten as
\begin{equation}\label{bp.109a}
{2^{1-l} \over 1 - \alpha^2 } \Big (
{\rm Pf} [ a_{jk} ]_{ 2(l+1) \times 2(l+1) } \Big |_{\rho_{2l+1} = \rho_{2l+2} = 0} -
(1 + \alpha^2) {\rm Pf} [ a_{jk} ]_{2l \times 2l} \Big ).
\end{equation}
Comparing (\ref{bp.106a}) and (\ref{bp.109}) tells us that
$$
 {\rm Pf} [ a_{jk} ]_{2l \times 2l} =
2^{l-1} \Big ( \alpha^{\sum_{j=1}^{2l} \rho_j {\rm mod} \, 2} +
 \alpha^{\sum_{j=1}^{2l} ( \rho_j + 1) {\rm mod} \, 2} \Big ).
$$
Substituting in (\ref{bp.109a}) and simplifying implies the matrix integral evaluation
\begin{equation}\label{bp.109d}
\langle \det ( {\bf 1}_{2l} + \alpha  U ) s_\rho( U) \rangle_{{ U} \in O^-(2l)} =
\alpha^{\sum_{j=1}^{2l} \rho_j {\rm mod} \, 2} - \alpha^{\sum_{j=1}^{2l}
(\rho_j + 1) {\rm mod} \, 2}.
\end{equation}
Adding this to (\ref{bp.109}) verifies (\ref{bp.OU}) in the case $l$ even.

Similar working suffices in the $l$ odd case, $l \mapsto 2l+1$. In regards to the
average over $O^+(2l+1)$, making use of the explicit form of the p.d.f.~(\ref{O.2}),
the determinant form of the Schur polynomial (\ref{sdet}), and integration techniques from
random matrix theory, one obtains the Pfaffian formula
\begin{equation}\label{dt.1}
\langle \det ({\bf 1}_{2l+1} + \alpha  U) s_\rho( U) \rangle_{{ U} \in O^+(2l+1)} =
{(1 + \alpha) \over 2^l}
{\rm Pf} \left [
\begin{array}{cc} [a_{jk}]_{j,k=1,\dots,2l+1} & [1]_{j=1,\dots,2l+1} \\
-[1]_{k=1,\dots,2l+1} & 0 \end{array} \right ]
\end{equation}
where $a_{jk}$ is specified by (\ref{5.13a}). This Pfaffian can in fact be
evaluated by making use of the Pfaffian evaluation implied by the equality of
(\ref{bp.106a}) and (\ref{bp.109}). To see this, multiply the final row and
column of the matrix in (\ref{dt.1}) by $(1+\alpha)^2$, and balance the equation
by dividing by a prrefactor of $(1+\alpha)^2$ on the r.h.s.. Next subtract the 2nd
last row from the final row, and 2nd last column from the final column. Finally,
write in the 2nd last entry of the final row and column $(1+\alpha)^2 = (1 + \alpha^2)
+ 2 \alpha$. This shows that (\ref{dt.1}) is equal to
\begin{equation}
{1 \over 2^l (1 + \alpha) }
{\rm Pf} [ A_{(2l+2) \times (2l+2)} \Big |_{\rho_{2l+2} = \rho_{2l+1}} + B']
\end{equation}
where $A_{(2l+2) \times (2l+2)} := [a_{jk}]_{j,k=1,\dots,2l+2}$ while $B'$ has all
entries zero except for the second last entry of the final column, which is $2 \alpha$,
and the second last entry of the final row, which is $-2 \alpha$. Making use of
(\ref{bp.107}) shows that this in turn is equal to
\begin{equation}\label{dt.2a}
{1 \over 2^l (1 + \alpha) } \Big (
{\rm Pf} \,  A_{(2l+2) \times (2l+2)} \Big |_{\rho_{2l+2}  = \rho_{2l+1}}
+ 2 \alpha {\rm Pf} \, A_{2l \times 2l} \Big ).
\end{equation}
But the value of ${\rm Pf} \, A_{2l \times 2l}$ for general $l$ is known from the equality
between (\ref{bp.106a}) and (\ref{bp.109}), so we find that (\ref{dt.2}) reduces to
\begin{equation}\label{dt.2}
\alpha^{\sum_{j=1}^{2l+1} \rho_j {\rm mod} \, 2} +
\alpha^{\sum_{j=1}^{2l+1} (\rho_j + 1) {\rm mod} \, 2}
\end{equation} 
thus giving the evaluation of the random matrix average in (\ref{dt.1}).

For the average over $O^-(2l+1)$, we note that a change of variables $\theta_j \mapsto \pi -
theta_j$ $(j=1,\dots,l)$ shows
\begin{equation}\label{dt.3}
\langle \det ({\bf 1}_{2l+1} + \alpha  U) s_\rho( U) \rangle_{{ U} \in O^-(2l+1)} =
(-1)^{|\rho|} \langle \det ({\bf 1}_{2l+1} - \alpha  U) s_\rho( U) \rangle_{{ U} \in O^+(2l+1)}.
\end{equation}
Substituting (\ref{dt.2}) with $\alpha \mapsto - \alpha$ for the average on the r.h.s.~shows that this
is equal to
\begin{equation}\label{dt.4}
\alpha^{\sum_{j=1}^{2l+1} \rho_j {\rm mod} \, 2} -
\alpha^{\sum_{j=1}^{2l+1} (\rho_j +1 ) {\rm mod} \, 2}. 
\end{equation}
Finally, taking the arithmetic mean of (\ref{dt.2}) and (\ref{dt.4}), we obtain the sought evaluation
(\ref{bp.OU}) with $l \mapsto 2l+1$. \hfill $\square$

\section{Matrices symmetric about both the diagonal and anti-diagonal}
\setcounter{equation}{0}
Let the $2n \times 2n$ matrix $X = [x_{i,j} ]_{i,j=1,\dots, 2n}$ have
non-negative integer entries, and label the rows from the bottom.
Suppose furthermore the entries are symmetric with respect to reflections
in both the diagonal ($x_{i,j} = x_{j,i},
\, i > j$) and anti-diagonal ($x_{i,j} = x_{i,2n+1-j}, \, i> 2n+1-j$).
Because $X$ is symmetric about the diagonal, the RSK correspondence maps
$X$ to a pair of identical non-intersecting u/rh lattice paths $(P,P)$
say. On the other hand $X$ being symmetric about the anti-diagonal implies
$X$ maps to the lattice path pair $(P^R,P)$, where $P^R$ is the
Sch\"utzenberger dual of $P$. Consequently in this case $X$ maps under the
RSK correspondence to a single set of at most $2n$ u/rh non-intersecting
lattice paths $P$ with the special property that $P=P^R$.

We will suppose furthermore that the entries on the anti-diagonal are
constrained to be even. Then according to (\ref{6.2r'}) all final
displacements $\mu_i$ 
of the paths must be even. A partition  with parts
$2 \lambda_i$ so each part is even 
will be written $2 \lambda$.

The independent elements of $X$ can be chosen to be $x_{i,j}$ with
$i \le j$ $(i,j=1,\dots,n)$ and $i \le 2n+1-j$
($i=1,\dots,n$, $j=n+1,\dots,2n$). We choose the value of each such
$x_{i,j}$, excluding those on the diagonal or anti-diagonal, according to
the geometric distribution
$$
{\rm Pr}(x_{i,j} = k) = (1 - q_i q_j) (q_i q_j)^k
$$
On the anti-diagonal we choose
\begin{equation}\label{dc.1}
{\rm Pr}(x_{i,2n+1-i}=k)  =  \left \{
\begin{array}{ll} (1 - q_i^2) q_i^k, \quad &  k \:\: {\rm even} \\
0, & {\rm otherwise} \end{array} \right.
\end{equation}
and on the diagonal
\begin{equation}\label{dc.2}
{\rm Pr}(x_{i,i} = k) = (1 - \alpha q_i) (\alpha q_i)^k \qquad
(i=1,\dots,n).
\end{equation}
The bijection then implies
that the probability $X$ maps to a set of at most $2n$
non-intersecting u/rh lattice paths of final displacement $2 \lambda$
is equal to
\begin{equation}\label{8.128a}
\prod_{i=1}^n\Big ((1 - q_i^2)(1 - \alpha q_i)  \prod_{j=i+1}^n(1 - q_i q_j)
\prod_{j=n+1}^{n+i}(1 - q_i q_j) \Big )
\alpha^{\sum_{l=1}^{2n} \lambda_l}
s_{2\lambda}^{\rm s.d.}(q_1,\dots,q_{2n})
\end{equation}
where
\begin{equation}\label{bp.ssu3}
s_{2\lambda}^{\rm s.d.}(q_1,\dots,q_{2n}) :=
\sum \nolimits^* 
\prod_{j=1}^{2n} q_j^{{1 \over 2} \sum_{l=1}^{2n}
\tilde{\lambda}_l(j)} 
\end{equation}
with $\tilde{\lambda}_l(j)$ denoting the number of vertical steps at $x=j-1$
contained in the level-$l$ path and the asterisk denoting that the sum is
over all self dual u/rh non-intersecting lattice paths with final displacement
$2 \lambda$. Because for such self dual lattice paths 
$\tilde{\lambda}_l(j) =
\tilde{\lambda}_l(2n+1-j)$ (recall the discussion below (\ref{5.sn})) we have
$$
\prod_{j=1}^{2n} q_j^{{1 \over 2} \sum_{l=1}^{2n}
\tilde{\lambda}_l(j)} 
= \prod_{j=1}^n (q_j q_{2n+1-j} )^{{1 \over 2} \sum_{l=1}^{2n}
\tilde{\lambda}_l(j)}.
$$
This allows us to set
$$
q_i = q_{2n+1-i} \qquad (i=1,\dots,n),
$$
and so with 
\begin{equation}\label{bp.114'}
\tilde{s}_{2\lambda}^{\rm s.d.}(q_1,\dots,q_{n}) :=
\sum \nolimits^*
s_{2\lambda}^{\rm s.d.}(q_1,\dots,q_{2n}) \Big |_{q_i = q_{2n+1-i}
\atop (i=1,\dots,n)} =
\prod_{j=1}^{n} q_j^{\sum_{l=1}^{2n}
\tilde{\lambda}_l(j)}
\end{equation}
(\ref{8.128a}) reads
\begin{equation}\label{8.128b}
\prod_{i=1}^n(1 - \alpha q_i)
\prod_{i,j=1}^n(1 - q_i q_j) \alpha^{\sum_{j=1}^{2n}
(-1)^{j-1}  \lambda_j}
\tilde{s}_{2\lambda}^{\rm s.d.}(q_1,\dots,q_{n}).
\end{equation}

\begin{figure}
\epsfxsize=8cm
\centerline{\epsfbox{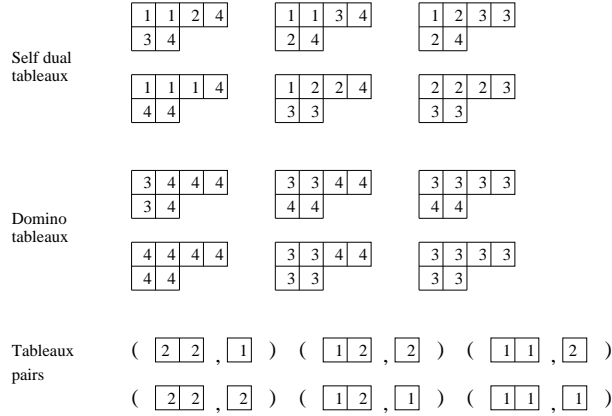}}
\caption{\label{bp.f22} The self dual tableaux of shape 42 and content 4,
the corresponding domino tableaux and the corresponding pair of
tableaux of shape 2, content 2, and shape 1, content 1. 
}
\end{figure}

As noted in \cite{BR01a}, the polynomial $\tilde{s}_{2\lambda}^{\rm s.d.}$ is
expressible in terms of Schur polynomials. To understand this point, one
must first establish a relation between self dual lattice paths, 
represented as self dual tableaux, and domino tableaux. Regarding the latter,
consider the diagram of a partition $2\lambda$. Define a domino tableau,
of shape $2\lambda$ with content from $\{n+1,\dots,2n\}$, 
as a tiling of the
diagram by dominos with the dominos numbered from the set
$\{n+1,\dots,2n\}$ (each number repeated twice to fill the two squares
of the domino) such that the number given to different dominos strictly
increase down columns and weakly increase along rows. It is a known result
(see e.g.~\cite{vL96}) that there is a bijection between self dual
tableaux of shape $2\lambda$, content $2n$, and domino tableaux 
of shape $2\lambda$ with content from $\{n+1,\dots,2n\}$. In particular,
to construct a domino tableau from a self dual tableau $P$ say, one applies
in succession the Sch\"utzenberger evacuation operation (see e.g.~\cite{Sa01}),
and the operation of removing the last square displaced in this operation.
The domino formed by the evacuated and removed squares is numbered
by the number of the removed square, which will be between $n+1$ and $2n$
(note that $P$ being self dual implies the sum of the entries of the
evacuated and removed squares is $2n+1$). The procedure is repeated until
all dominos have been identified and numbered (see Figure \ref{bp.f22}
for an example of the end product of this mapping).

Having established the bijection between self dual tableaux and domino
tableaux, one now makes use of a bijection  between domino tableaux of
shape $2\lambda$ and content from $\{n+1,\dots,2n\}$, and pairs of
semi-standard tableaux $(P,Q)$ of shape $(\mu,\kappa)$ each of
content $n$ with
\begin{equation}\label{bp.rq}
{\rm shape}\, P = (\lambda_1,\lambda_3,\dots,\lambda_{m})
=: \lambda^+, \qquad
{\rm shape}\, Q = (\lambda_2,\lambda_4,\dots,\lambda_{m-1}) =: \lambda^-
\end{equation}
where $m$ equals the length of $2\lambda$ if the latter is odd, and
one minus the length if it is even (and then $\lambda_m=0$). To construct
$P$ ($Q$), remove all columns from the domino tableau for which the
absolute value of
the column number minus the length of the  column is odd (even), and remove
all even (odd) numbered rows. Finally subtract $n$ from each of the
numbers (an example of the result of this mapping is given in
Figure \ref{bp.f22}). Recalling the definition (\ref{bp.114'}) of
$\tilde{s}_{2\lambda}^{\rm s.d.}$ and the definition (\ref{3.Sc}) of the
Schur polynomials, it follows that
\begin{equation}\label{bp.rqa}
\tilde{s}_{2\lambda}^{\rm s.d.}(q_1,\dots,q_n) =
s_{\lambda^+}(q_1,\dots,q_n)
s_{\lambda^-}(q_1,\dots,q_n).
\end{equation}
Consequently the probability (\ref{8.128b}) can be written as
\begin{equation}\label{8.128c}
\prod_{i=1}^n(1 - \alpha q_i)
\prod_{i,j=1}^n(1 - q_i q_j)\alpha^{\sum_{j=1}^{2n}(-1)^{j-1}
\lambda_j}
s_{\lambda^+}(q_1,\dots,q_n)
s_{\lambda^-}(q_1,\dots,q_n).
\end{equation}
In the recent work \cite{FR02b}
(\ref{8.128c}) was stated without derivation as being equal to the
probability that the $2n \times 2n$ matrix $X$, symmetric about both the
diagonal and anti-diagonal, and with elements distributed according to
(\ref{dc.1}) and (\ref{dc.2}), maps under the RSK correspondence to a set
of at most $2n$ non-intersecting u/rh lattice paths of final displacement
$2 \lambda$. This is precisely the result derived here.

From the definition (\ref{3.Sc}) of the Schur polynomials in terms of 
non-intersecting lattice paths, it is easy to see that the well known
identity
$$
\sum_{\lambda^- : \: \lambda^+ \: {\rm fixed}}
\alpha^{\sum_{j=1}^{2n}(-1)^{j-1}
\lambda_j} s_{\lambda^-}(q_1,\dots,q_n) =
s_{\lambda^+}(q_1,\dots,q_n,\alpha)
$$
holds. Thus the marginal probability of $\lambda^+$ in (\ref{8.128c})
is equal to
\begin{equation}\label{6.1p}
\prod_{i=1}^n (1 - \alpha q_i)
\prod_{i,j=1}^n (1 - q_i q_j)
s_{\lambda^+}(q_1,\dots,q_n) s_{\lambda^+}(q_1,\dots,q_n,\alpha).
\end{equation} 
This in turn implies that with
$$
L^\symmu_{2n} := {\rm max} \sum_{(1,1){\rm u/rh}(2n,2n) \atop X=X^T=X^R} x_{i,j}
$$
we have
\begin{eqnarray}\label{6.1pa}
&&
{\rm Pr}(L^\symmu_{2n} \le 2l) = {\rm Pr}(L^\symmu_{2n} \le 2l+1) =
\prod_{i=1}^n(1 - \alpha q_i) \prod_{i,j=1}^n(1 - q_i q_j) \nonumber \\&&
\qquad \times \sum_{\lambda^+: \lambda_1 \le l}
s_{\lambda^+}(q_1,\dots,q_n) s_{\lambda^+}(q_1,\dots,q_n,\alpha).
\end{eqnarray} 
The sum in (\ref{6.1pa}) is a special case of that in (\ref{6.2}) ---
thus replace $n \mapsto n+1$ in the latter and set
$a_i = b_i = q_i$ $(i=1,\dots,n)$, $a_{n+1}=0$, $b_{n+1}=\alpha$. It
therefore follows from (\ref{3.1}) that
\begin{eqnarray}
&&
{\rm Pr}(L^\symmu_{2n}\le 2l) = {\rm Pr}(L^\symmu_{2n} \le 2l+1) \nonumber \\&&
\qquad  =
\prod_{i=1}^n(1 - \alpha q_i) \prod_{i,j=1}^n(1 - q_i q_j)
\Big \langle \prod_{k=1}^l (1+\alpha e^{i  \theta_k})
\prod_{j=1}^n \prod_{k=1}^l
|1 + q_j e^{i \theta_k}|^2
\Big \rangle_{U(l)}
\end{eqnarray}
This result is the special case $\beta = 0$ (the effect of setting
$\beta = 0$ is to constrain the elements on the anti-diagonal of $X$
to be even) 
of a result first derived in \cite{BR01a} using methods of symmetric
function theory to sum over a 
$\beta$-generalization of (\ref{8.128a}). 

\section{Matrices with a point reflection symmetry}
\setcounter{equation}{0}
The point $(n+1/2,n+1/2)$ is at the centre of the region $1 \le x,y \le 2n$.
A point $(x,y)$ in this region reflected about this central point maps to
the point $(2n+1-x,2n+1-y)$. We thus say that a $2n \times 2n$ matrix $X$
has a point reflection symmetry (about the point $(n+1/2,n+1/2)$) if
$x_{i,j} = x_{2n+1-i,2n+1-j}$ $(i,j=1,\dots,n)$ or consequently
if $X = (X^R)^T$. For a matrix with this symmetry we can take as the
independent elements the triangular region below the anti-diagonal $i < 2n+1-j$ ($i,j=1,\dots,
2n$) together with the portion of the anti-diagonal
$i= 2n+1-j$ ($i,j=1,\dots,n$). 

We seek the constraint on the pairs of paths $(P_1,P_2)$, with both $P_1$
and $P_2$ of the same final displacements, which according to the RSK
mapping are in correspondence with matrices $X$ with a point reflection
symmetry. We have already noted that with $X$ mapping under RSK to
$(P_1,P_2)$, $X^R := [ x_{2n+1-j,2n+1-i}]_{i,j=1,\dots,2n}$ maps to
$(P_2^R,P_1^R)$ while $X^T := [x_{j,i}]_{i,j=1,\dots,2n}$ maps to
$(P_1^R,P_2^R)$ and hence matrices with the point reflection symmetry
$X = (X^R)^T $ map to a pair of u/rh lattice paths of the same final
displacements constrained so that
\begin{equation}
P_1 = P_1^R, \qquad P_2 = P_2^R .
\end{equation}

We choose the independent entries of $X$ according to the geometric
distribution
$$
{\rm Pr}(x_{i,j} = k) = (1 - q_i q_j) (q_i q_j)^k
$$
where to be compatible with the point reflection symmetry we require
$q_{2n+1-i} = q_i$ $(i=1,\dots,n)$. With this specification it follows
from the bijection
that the probability $X$ maps to a pair of u/rh lattice paths of final
displacement $\mu$ is equal to
$$
\Big ( \prod_{i,j=1}^n(1 - q_i q_j)
\tilde{s}_{\mu}^{\rm s.d.}(q_1,\dots,q_n) \Big )^2.
$$

We know from (\ref{bp.rqa}) that when the length of the parts of $\mu$
are all even, $\tilde{s}_{\mu}^{\rm s.d.}$ can be given in terms of
Schur polynomials. This is also true in the general case \cite{vL96}. One
again proceeds by noting that there is a bijection between a general
self dual tableaux of shape $\lambda$ and domino tableaux. 
An immediate consequence is that unless $\lambda$ admits a domino tiling
--- for which the necessary and sufficient condition is that
the number of points
$(i,j)$ in the diagram of $\lambda$
with $i+j$ even is equal to the number of points
with $i+j$ odd --- one has $\tilde{s}_{\mu}^{\rm s.d.} = 0$. 
It is also true that
domino tableaux are in bijective correspondence with pairs of
semi-standard tableaux of shape $(\mu^{(0)}, \mu^{(1)})$,
$|\mu^{(0)}| + |\mu^{(1)}| = |\mu|$, each of content $n$ where
$\mu^{(0)}$ and $\mu^{(1)}$ are the so called 2-quotient of the partition
$\mu$. Regarding the latter, let $\mu$ be a partition of length $m$.
Add to $\mu$ the  partition 
$\delta_m := (m-1,m-2,\dots,1,0)$, and from this construct two new
partitions $\tilde{\mu}^{(0)}$, $\tilde{\mu}^{(1)}$ of lengths
$m^{(0)}, m^{(1)}$, the first consisting of the even parts of
$\mu+ \delta_m$, and the  second the odd parts of $\mu+ \delta_m$. 
The 2-quotient is the pair of partitions $\mu^{(0)}$, $\mu^{(1)}$
specified by \cite{Ma95}
$$
\mu^{(0)} = \tilde{\mu}^{(0)}/2 - \delta_{m^{(0)}}, \qquad
\mu^{(1)} = (\tilde{\mu}^{(1)}+1)/2 - \delta_{m^{(1)}}.
$$

Analogous to (\ref{bp.rqa}) one therefore has that if $\mu$ admits a
domino tiling, then
\begin{equation}\label{8.134}
\tilde{s}_{\mu}^{\rm s.d.}(q_1,\dots,q_n) =
\tilde{s}_{\mu^{(0)}}(q_1,\dots,q_n)
\tilde{s}_{\mu^{(1)}}(q_1,\dots,q_n). 
\end{equation}
Regarding the converse of this statement, it's easy to see that the 2-quotient
of a partition which admits a domino tiling is unique, while the 
2-quotient of a partition which does not admit a domino tiling
coincides with the 2-quotient of a partition which does. Hence, given
arbitrary partitions $\mu^{(0)}$, $\mu^{(1)}$ there is a unique
partition $\mu$ which admits a domino tiling and is such that
(\ref{8.134}) is satisfied. Furthermore, from the definition of a 
2-quotient, if $\mu_1 \le 2l$ then $\mu^{(0)}_1 \le l$ and 
$\mu^{(1)}_1 \le l$,
while if $\mu_1 \le 2l+1$ then $\mu^{(0)}_1 \le l+1$ and $\mu^{(1)}_1 \le l$,
or $\mu^{(1)}_1 \le l+1$ and $\mu^{(0)}_1 \le l$. Thus with
$$
L^{\symmUU}_{2n} := {\rm max} \sum_{(1,1) {\rm u/rh} (2n,2n)} x_{i,j}
$$
we have
\begin{eqnarray}\label{9.1}
&&
{\rm Pr}(L^{\symmUU}_{2n} \le 2l) = \prod_{i,j=1}^n (1 - q_i q_j)^2
\sum_{\mu: \mu_1 \le 2l}
\Big (\tilde{s}_{\mu}^{\rm s.d.}(q_1,\dots,q_n) \Big )^2
\nonumber \\
&& \qquad =
\Big ( \prod_{i,j=1}^n(1 - q_i q_j) \sum_{\kappa: \kappa_1 \le l}
(s_\kappa(q_1,\dots,q_n))^2 \Big )^2 =
\Big ( {\rm Pr}(L^{\symmU}_n \le l) \Big )^2
\Big |_{\{a_i\} = \{b_i\}=\{q_i\}}
\end{eqnarray}
where the final equality follows upon comparison with (\ref{6.2}), and
\begin{eqnarray}\label{9.2}
&&
{\rm Pr}(L^{\symmUU}_{2n} \le 2l+1) = \prod_{i,j=1}^n (1 - q_i q_j)^2
\sum_{\mu: \mu_1 \le 2l+1}
\Big (\tilde{s}_{\mu}^{\rm s.d.}(q_1,\dots,q_n) \Big )^2
\nonumber \\
&& \qquad =
\prod_{i,j=1}^n(1 - q_i q_j)^2 \sum_{\kappa: \kappa_1 \le l+1}
(s_\kappa(q_1,\dots,q_n))^2 
\sum_{\kappa: \kappa_1 \le l}
(s_\kappa(q_1,\dots,q_n))^2 \nonumber \\&&
 \qquad =
 {\rm Pr}(L^{\symmU}_n \le l+1) 
\Big |_{\{a_i\} = \{b_i\}=\{q_i\}}
{\rm Pr}(L^{\symmU}_n \le l) 
\Big |_{\{a_i\} = \{b_i\}=\{q_i\}}.
\end{eqnarray}
The results (\ref{9.1}) and (\ref{9.2}) were stated without derivation
in \cite{BR01}. The derivation given here uses the strategy outlined
in \cite{BR01a} to derive (\ref{9.1}) and (\ref{9.2}) in the exponential
limit which corresponds to
the appropriately symmetrized form of the Hammersley process. 


\end{document}